\documentclass[twocolumn,aps]{revtex4}
\usepackage{epsfig}
\usepackage{psfig}
\usepackage{amsmath}

\newcommand{\be}{\begin{equation}}
\newcommand{\ee}{\end{equation}}
\newcommand{\bea}{\begin{eqnarray}}
\newcommand{\eea}{\end{eqnarray}}

\begin{document}
\title{Could quantum gravity be tested with high intensity Lasers?}
\author{Jo\~{a}o Magueijo$^{1,2,3}$ }
\affiliation{ $^1$Perimeter Institute for Theoretical Physics, 31
Caroline St N, Waterloo N2L 2Y5, Canada\\
$^2$ Canadian Institute for Theoretical Astrophysics,
60 St George St, Toronto M5S 3H8, Canada\\
$^3$ Theoretical Physics Group, Imperial College, Prince Consort
Road, London SW7 2BZ, England}

\begin{abstract}
In quantum gravity theories Planckian behavior is triggered by the
energy of {\it elementary} particles approaching the Planck
energy, $E_P$, but it's also possible that anomalous behavior
strikes systems of particles with total energy near $E_P$. This is
usually perceived to be pathological and has been labelled ``the
soccer ball problem''. We point out that there is no obvious
contradiction with experiment if {\it coherent} collections of
particles with bulk energy of order $E_P$ do indeed display
Planckian behavior, a possibility that would open a new
experimental window. Unfortunately field theory realizations of
deformed special relativity never exhibit a ``soccer ball
problem''; we present several formulations where this is
undeniably true. Upon closer scrutiny we discover that the only
chance for Planckian behavior to be triggered by large coherent
energies involves the details of second quantization. We find a
formulation where the quanta have their energy-momentum
(mass-shell) relations deformed as a function of the bulk energy
of the coherent packet to which they belong, rather than the
frequency. Given ongoing developments in Laser technology, such a
possibility would be of great experimental interest.
\end{abstract}
\pacs{PACS Numbers: *** }
\keywords{
}
\date{\today}

\maketitle

\section{Introduction}
Non-linear or deformed special relativity (hereafter DSR) is an
interesting arena for studying quantum gravity (QG) phenomenology
at energies close to the Planck energy $E_P=1/l_P\approx
10^{19}$~Gev (\cite{amel1,ljprl,leej,desitter}). Typically in such
theories the Planck energy is an invariant and particles can never
exceed this energy. For example a photon cannot be blue-shifted
over the Planck energy; likewise an hypothetical photon with
energy $E_P$ would be seen with $E_P$ by all observers. These
desirable properties ensure an invariant separation between the
realms of classical and quantum gravity and provide an invariant
cut off removing divergences from field theory.

But they also entail  apparent paradoxes~\cite{soc1,soc2}. Most
infamously a naive application of the formalism leads to the
conclusion that not only do single particles have a maximal energy
$E_P$, but this property extends to collections of particles.
Systems of particles, notably soccer balls, can blatantly exceed
$E_P$; hence the eponymous problem.

There are a number of ad-hoc solutions to the ``soccer ball
problem'' (see, eg.~\cite{leej,judes}). One approach posits that a
system of particles simply does not have the same transformation
laws and dispersion relations as its constituents~\cite{leej}, and
that the distinction between elementary  and composite is
fundamental. Specifically it was proposed that a system of $N$
particles obeys non-linear transformations and non-quadratic
dispersion relations obtained by replacing $E_P$ by $NE_P$ (or
$l_P$  by $l_P/N$). This implies a non-associative addition law
for energy and momentum, since one cannot ``associate'' terms in a
sum thereby losing track of how many elementary particles it
contains. The maximal energy for a system of $N$ particles is now
$NE_P$, resolving the paradox.

Solving the soccer-ball problem ensures consistency with
experiment, but also limits testability. In this paper we
investigate whether there might be a middle ground where
collections of particles retain the imprint of QG in a way that is
not obviously inconsistent with experiment. For example, it could
it be that loss of coherence is essential for solving the soccer
ball problem. Perhaps a coherent superposition of particles (e.g.
a laser) does have maximal energy $E_P$, and QG effects are of the
order of the ratio between its {\it total} energy and $E_P$. The
macroscopic objects we commune with in daily life are
non-coherent, and perhaps only these are protected from soccer
ball anomalies.

This speculation is not immediately at odds with experiment. Laser
beams are the most powerful coherent superpositions of particles
available, but the largest energy attained so far is about
$10^{12}$ Gev, much smaller than $E_P$~\cite{nova,nif}. Laser
technology is improving fast, however, so the possibility that a
laser carrying a bulk energy close to $E_P$ might display strong
QG behavior suggests a remarkable competitor to ultra high energy
cosmic rays (UHECRs). These are widely regarded as {\it the} test
bed of QG and DSR, because the predicted GZK energy cut
off~\cite{gzk} probes Lorentz boosts at the highest available
energies~\cite{amelgzk}. If the cutoff at $E\approx 10^{11}$~Gev
predicted by special relativity is present, however, then high
energy cosmic rays will dry out as a probe of QG. Not so with
Lasers. The currently operational NOVA laser~\cite{nova} can
deliver $400$ KJ of coherent light, i.e. $2.5\times 10^{12}$ Gev,
already larger than the highest cosmic rays detected. The planned
National Ignition Facility~\cite{nif} will raise this figure to
$2.5\times 10^{13}$ Gev.

Regrettably this possibility is by no means a generic feature of
DSR or QG. The arguments leading to the soccer-ball problem are
purely kinematic and refer to classical point particles. In this
paper we reevaluate the situation from a field theory perspective.
In field theory realizations of DSR there should be classical
plane wave solutions with wavenumber $k_\mu=(\omega,-{\mathbf
k})$, constrained by deformed dispersion relations (with a
deformation dependent on $l_P\omega$, say). Their amplitude $A$
further tunes the bulk energy ${\cal E}$ of the wave. These waves
form our ``soccer balls''. Upon quantization the quanta have
energy and momentum proportional to $\omega$ and ${\mathbf k}$,
and therefore feel dispersion relations deformed according to
$l_P\omega$. It does not follow that the wave's bulk energy ${\cal
E}$ and momentum ${\cal P}_i$, dependent on amplitude $A$, must
feel deformations according to $l_P{\cal E}$ as opposed to
$l_P\omega$. Indeed we show that it is very difficult to have a
soccer ball problem in a field theory formulation of DSR.

The plan of the paper is as follows.  In Section~\ref{model} we
carry out, in the undeformed theory, the model calculation  later
to be performed in field theories representing DSR. We use the
stress-energy tensor to evaluate the mass-shell relations for the
wave's bulk  ${\cal E}$ and ${\cal P}_i$, and contrast them with
the dispersion relations for $k_\mu$. Then, in Section~\ref{field}
we explain how DSR may be represented by higher order derivative
(HOD) field theories. Although we have Lasers in mind we keep our
arguments general and consider massless and massive particles. For
massless particles we consider the possibility of an energy
dependent speed of light~\cite{amelvsl,vslrev}, but this is by no
means necessary. For simplicity we ignore spin and examine a real
scalar field, but the constructions presented may be easily
generalized to any spin including spin 1. We set up a Lagrangian
formulation for HOD theories and derive their stress energy
tensor, essential for the assessment of the macroscopic properties
of the waves.

Equipped with these tools, in Section~\ref{emwaves} we examine two
concrete examples of DSR theories, where $k_\mu$ has  deformed
transformation laws and dispersion relations. We find that the
deformations felt by bulk quantities ${\cal E}$ and ${\cal P}_i$
depend on $l_P\omega$, not $l_P{\cal E}$. Thus, as long as
$\omega$ remains well below $E_P$ the deformations felt by the
bulk wave are very small, regardless of the total ${\cal E}$, and
therefore there is no soccer ball problem. The argument does not
rely on the details of second quantization, however in
Section~\ref{quantize} we show how it is possible to quantize in
such a way that the dispersion relations felt by a system of $N$
particles can be obtained by the replacement $l_P\rightarrow
l_P/N$, thus proving the suggestion in~\cite{leej} from first
principles.

This conclusion exempts DSR models from a paradox, but does not
satisfy our motivations. We therefore spend the rest of the paper
seeking theories that might have phenomena akin to the soccer ball
problem,  preferably only in the case of coherent objects.
Non-quadratic Lagrangian theories are considered in
Section~\ref{NQL}. They have the property that both quanta and
bulk dispersion relations are dependent on the energy density,
i.e. on ${\cal E}/(V E_P^4)$. Therefore they also don't display a
soccer ball problem. In Section~\ref{laser} we argue that this
conclusion is unavoidable without appealing to an unorthodox
second quantization. This is developed in Section~\ref{abnormal},
where an exotic second quantization procedure is presented (to be
contrasted with that in Section~\ref{quantize}), for which the
quanta's mass-shell conditions are deformed according to $l_P{\cal
E}$, where ${\cal E}$ is the energy of the coherent packet to
which the quanta belong. This is undoubtedly unusual, but cannot
be ruled out a priori. Thus we satisfy our experimental
motivation. An overview is provided in the concluding Section.

Throughout this paper we use a metric with signature $+ - - -$.
We choose units where {\it in the low energy limit},
$\hbar=c=G=1$. Whenever we consider frequency dependent functions
$\hbar(\omega)$ or
$c(\omega)$ these refer to dimensionless ratios
with the low energy values of $\hbar$ or $c$.

\section{A model calculation}\label{model}
The basic idea behind this paper is that plane waves (or wave
packets) form excellent test tubes with which to probe the
behavior of sets of coherent particles in deformed special
relativity. We can use their amplitude to tune the number $N$ of
particles they contain. We can then compute the stress-energy of
the wave seen as a bulk and check directly what the dispersion
relations are for a set of $N$ coherent particles. Such coherent
superpositions of quantum particles in effect form classical waves
and provide good models for lasers.

We start by illustrating what we're hoping to do by considering
the undeformed theory, taking as a prototype a free
field theory with Lagrangian
\be \label{LagKG0}{\cal L}={1\over
2}\left[\partial_\mu\phi\partial^\mu\phi - m^2\phi^2\right] \,
.\ee
As explained in the introduction we shall ignore the complications
of spin (polarization), but the argument can easily be
generalized.  Variation with respect to $\phi$ leads to the
Klein-Gordon equation
\be \label{KG0}[\partial_\mu\partial^\mu + m^2]\phi=0 \ee
which accepts plane wave solutions%
\be \label{pw}\phi=A e^{-ix^\mu k_\mu}\ee%
or, more precisely %
\be%
\label{pw1}\phi=A \cos(x^\mu k_\mu)\ee%
(we can only use the complex notation for the real field $\phi$ as
long as we are dealing with linear operations). The quanta, or
particles, in this theory have momentum
$k_\mu=(\omega,k_i)= (\omega,-k^i)$ and satisfy quadratic dispersion relations%
\be \omega^2-{\mathbf k}^2=m^2.\ee%
The proposed exercise is now to work out the dispersion relations for a
collection of these quanta, by evaluating the energy ${\cal E}$
and momentum ${\cal P}_i$ of the wave~(\ref{pw}).

Noether's theorem tells us that we need to collect the surface
terms (or full divergences) eliminated in going from
(\ref{LagKG0}) to (\ref{KG0}) in order to identify the conserved
currents associated with symmetries of the action. Specifically,
translations $x^\mu \rightarrow x^\mu +\delta x^\mu$ induce field
variations $\delta \phi=\partial_\mu \phi \delta x^\mu$. These in
turn induce a variation in the action
given only by the full divergence%
\be \delta S=\int dx^4 \partial_\mu {\left( {\partial{\cal L}\over
\partial(\partial_\mu\phi)}\delta\phi\right)}=
\int dx^4 \partial_\mu {\left( {\partial{\cal L}\over
\partial(\partial_\mu\phi)}\partial_\nu\phi\right)}\delta x^\nu\nonumber%
\ee%
assuming that the field equations
are satisfied.
But the action does not depend explicitly on $x^\mu$ so another
way to obtain this variation is to replace the fields by their
explicit functions of $x^\mu$ in order to identify ${\cal L}(x)=
{\cal L}(\phi(x),\partial_\mu\phi(x))$ and then compute%
\be%
\delta S=\int dx^4 {\partial{\cal L}\over \partial x^\nu}\delta x^\nu
=\int dx^4 \partial_\mu(g^\mu_\nu {\cal L})\delta x^\nu
.\ee%
Comparing the two expressions we thus obtain the result that%
\be%
T_{\mu\nu}={\partial{\cal L}\over
\partial(\partial^\mu\phi)}\partial_\nu \phi-g_{\mu\nu}{\cal
L}\ee%
is divergence free, that is
$\partial_\mu T^{\mu\nu}=0$.

For theory (\ref{LagKG0}) the stress-energy
tensor is:%
\be T_{\mu\nu}=
\partial_\mu\phi\partial_\nu\phi- g_{\mu\nu} {\cal
L}.\ee
Applying this expression to a wave we get%
\bea%
T_{00}&=&{A^2\over2}(\omega^2-{\mathbf k}^2\cos(2x^\mu k_\mu))\\
T_{0i}&=&{A^2\over 2}\omega k_i (1-\cos(2 x^\mu k_\mu))\eea%
(we should use the real expression (\ref{pw1}) here rather than
(\ref{pw}) since this operation is non-linear). Integrating over a
sufficiently large volume (with respect to the wavelength) the
oscillatory terms in these expressions vanish. Thus the wave's
bulk energy ${\cal E}$ and
momentum ${\cal P}_i$ inside a volume $V$ is given by%
\bea {\cal E}&=&T_{00}V={1\over 2} A^2 \omega^2 V\\
{\cal P}_i&=&T_{0i} V={1\over 2} A^2 \omega k_i V. \eea %
Given that the wave's energy and momentum is the sum of its
quanta's energy $\omega$ and momentum $k_i$, we learn that
the number of quanta contained in the wave is%
\be\label{number0} N={1\over 2}V A^2\omega\ee%
so that%
\bea%
{\cal E}&=&N\omega\label{nom}\\
{\cal P}_i&=&N k_i\label{nk}.%
\eea%
The basic assumption is that there is additivity when going from
the (quantum) parts to the (classical, coherent) whole.

We can now check explicitly that the set of particles, i.e. the wave
seen as a bulk, also satisfies quadratic
dispersion relations,
\be {\cal E}^2 -{\cal P}^2=N^2(\omega^2-k^2)=(Nm)^2 \ee%
The mass of the bulk equals the sum of masses of its quanta. Even
though in this case we have considered a coherent superposition we
see that the issue of coherence only affects the counting of $N$
in the wave, i.e. whether the amplitudes add or simply their
squares.

None of this is surprising but it should be obvious that several
steps in the calculation break down under deformed dispersion
relations. We'll now try to repeat this ``model calculation'' in
field theories representing deformed special relativity, in the
hope of identifying dispersion relations for (coherent)
collections of quanta. This should illuminate the soccer ball
problem from a more fundamental perspective.


\section{Higher-order field theories}\label{field}
Before that, however, we need to develop some tools. A possible
method for introducing deformed dispersion relations into field
theory appeals to Lagrangians with higher order derivatives
(see~\cite{dag}). We'll need to generalize the methods for
obtaining the Euler-Lagrange equation and the stress energy tensor
for such theories. We stress that in what follows the variables
$x^\mu$ are to be seen as standard (commutative) variables, and
that this is to be contrasted with approaches based on
non-commutative geometry~\cite{lukie,majid,kappa}. The issue of
invariance (and consequent energy dependence of the metric) is
discussed in \cite{dag} and \cite{rainbow}, and can be ignored at
this level.

Consider an action where the Lagrangian depends on higher order
derivatives%
\be S=\int dx^4 {\cal L} (\phi,\partial_\mu\phi,
\partial_{\mu\nu}\phi, \partial_{\mu\nu\alpha}\phi \cdots).
\ee%
Variation with respect to $\phi$ now leads to a much more complex
structure of surface terms, since we have to convert terms of the
form
\be
{\partial {\cal L}\over \partial(\partial_{\mu\nu...\beta}\phi)}
\partial_{\mu\nu...\beta}\delta\phi
\ee
into terms proportional to $\delta\phi$. For a term involving
$n$ derivatives of $\delta\phi$
this can be achieved by integrating by parts $n$ times, from which
result $n$ full divergences. The final result entails the
generalized Euler-Lagrange equations%
\be \label{EL}{\partial {\cal L}\over \partial\phi} -\partial_\mu
{\partial {\cal L}\over \partial(\partial_\mu\phi)} +
\partial_{\mu\nu} {\partial {\cal L}\over
\partial(\partial_{\mu\nu}\phi)} -
\partial_{\mu\nu\alpha}
{\partial {\cal L}\over \partial(\partial_{\mu\nu\alpha}\phi)} +
\cdots =0 \ee%
where we note that the sign alternates depending on the order of the
derivative in each term. The divergences generated by this process
have the  form
$\partial_\mu D^\mu$ with %
\bea \label{divs}D_\mu&=&{\partial {\cal L}\over
\partial(\partial^{\mu}\phi)}\delta\phi + \nonumber\\
&&{\partial {\cal L}\over
\partial(\partial^{\mu}_\alpha\phi)}\partial_{\alpha}\delta\phi
-\partial_\alpha{\partial {\cal L}\over
\partial(\partial^{\mu}_\alpha\phi)}\delta\phi +\nonumber\\
&&{\partial {\cal L}\over
\partial(\partial^{\mu}_{\alpha\beta}\phi)}\partial_{\alpha\beta}\delta\phi
-\partial_\alpha{\partial {\cal L}\over
\partial(\partial^{\mu}_{\alpha\beta}\phi)}\partial_{\beta}\delta\phi
+\partial_{\alpha\beta}{\partial {\cal L}\over
\partial(\partial^{\mu}_{\alpha\beta}\phi)}\delta\phi
\nonumber\\
&&\cdots\nonumber \eea%
Following argument identical to the one in the previous section we
can therefore identify
the conserved stress-energy tensor%
\bea \label{Tmn}T_{\mu\nu}&&={\partial {\cal L}\over
\partial(\partial^{\mu}\phi)}\partial_\nu\phi+\nonumber\\
&&{\partial {\cal L}\over
\partial(\partial^{\mu}_\alpha\phi)}\partial_{\alpha\nu}\phi
-\partial_\alpha{\partial {\cal L}\over
\partial(\partial^{\mu}_\alpha\phi)}\partial_{\nu}\phi+\nonumber\\
&&{\partial {\cal L}\over
\partial(\partial^{\mu}_{\alpha\beta}\phi)}\partial_{\alpha\beta\nu}\phi
-\partial_\alpha{\partial {\cal L}\over
\partial(\partial^{\mu}_{\alpha\beta}\phi)}\partial_{\nu\beta}\phi
+\partial_{\alpha\beta}{\partial {\cal L}\over
\partial(\partial^{\mu}_{\alpha\beta}\phi)}\partial_{\nu}\phi
\nonumber\\
&&\cdots\nonumber\\
&&-g_{\mu\nu}{\cal L}\eea%
It can be checked directly that the condition
\be\label{consTmn}%
\partial_\mu T^{\mu\nu}=0\ee%
is equivalent to the Euler-Lagrange equation (\ref{EL}). Since the
position variables $x^\mu$ remain standard commutative variables,
the concepts of translational symmetry and its associated
conserved stress energy tensor remain unmodified. In particular
the law (\ref{consTmn}) remains the same, except that the
expression for $T_{\mu\nu}$ is much more complex. But the only
novelties in this section are of a technical nature.

Higher order derivative (HOD) field theories may be used to
represent modified dispersion relations and by extension
DSR~\cite{dag,sab}. The prescription is that the field equation
should be obtained from the replacement \be \label{presc}
k_a\rightarrow i\partial_a \ee applied to whatever deformed
dispersion relation, \be \label{disp} \omega ^2 f^2 (\omega) -
{\mathbf k}^2 g^2(\omega) = m^2 \ee one wants to represent. Of
course we may algebraically rearrange the dispersion relations
before applying this prescription, thus leading to different field
equations. This is the same ambiguity associated with the fact
that the same dispersion relation may be represented by a variety
of non-linear representations of the Lorentz group~\cite{leej}. In
practice the particular representation chosen fully fixes the
field equation used.

For example, if the proposed particle kinematics is given by the
invariant: %
\be\label{disp1} {\omega^2-{\mathbf k}^2\over 1- (l_P
\omega)^2}=m^2 \ee %
the field equation should be:%
\be \label{kg1}\left[{\partial_\mu\partial^\mu\over
1+(l_P\partial_0)^2}+m^2\right]\phi=0\ee %
This equation is higher than second order, is linear (i.e. accepts
a superposition principle) and has plane wave solutions that
satisfy dispersion relations (\ref{disp1}). A less trivial matter
is finding a Lagrangian from which (\ref{kg1}) can be derived. One
possibility is \be {\cal L}=-{1\over
2}\phi{\left[{\partial_\mu\partial^\mu\over
1+(l_P\partial_0)^2}+m^2 \right]} \phi .\ee%
More generally ${\cal L}$ can be obtained by sandwiching the
deformed Klein-Gordon operator between two fields, but only if the
dispersion relations contain no odd powers of $\omega$. This is
equivalent to demanding that functions $f$ and $g$ in (\ref{disp})
are functions of $\omega^2$, or that their expansions in powers of
$l_P\omega$ only has even powers. The dispersion relations are
then symmetric under $\omega\rightarrow -\omega$, that is positive
and negative frequencies (energies) are treated equally.

In terms of the field theory this requirement means that the
Klein-Gordon operator should be real and only contain even order
derivatives. If one begins with a dispersion relation which has
odd powers of $l_P\omega$ one may still construct a Klein-Gordon
operator and a Lagrangian as before; however the field equation
derived from it will automatically be symmetrized, as will the
dispersion relations it represents. We shall therefore assume that
the dispersion relations have $\omega\rightarrow -\omega$
symmetry, even though this excludes some outstanding
examples~\cite{kappa,ljprl}.

As explained above, one may propose many different field theories
corresponding to the same dispersion relations (but there is a
one-to-one relation with the particular DSR or  non-linear
representation of the Lorentz group chosen). For example the
expression (\ref{disp1}) may be arranged as $\omega^2-{\mathbf
k}^2=m^2( 1- (l_P \omega)^2)$. Applying prescription (\ref{presc})
to these two equivalent expressions leads to inequivalent field
theories; the latter is no more than a (linear, non-frequency
dependent) redefinition of the units of frequency (or energy).

\section{The energy and momentum of waves}\label{emwaves}
Let us take as an example dispersion relations \be\label{disp11}
{\omega^2-{\mathbf k}^2\over 1- (l_P
\omega)^4}=m^2 \ee %
which we assume belong to a non-linear representation of the
Lorentz group with generators $K_i=U^{-1} L_{0i} U$, where
$L_{0i}$ are
the usual linear generators, and %
\be%
U(\omega,k_i)=(\omega(1+m^2l_P^4\omega^2)^{1/2},k_i)\ee%
(the procedure is described in detail in \cite{leej}). Then, we
should rewrite the dispersion relations as \be\label{disp11b}
\omega^2(1+m^2l_P^4\omega^2)- {\mathbf k}^2=m^2, \ee before
applying prescription (\ref{presc}). This leads to the field
equation \be (\partial_\mu\partial^\mu +m^2
-l_P^4m^2\partial_0^4)\phi=0 \ee which indeed has plane wave
solutions with a $k_\mu$ satisfying dispersion relations
(\ref{disp11}). A possible Lagrangian (giving this equation via
(\ref{EL})) is \be {\cal L}={1\over 2}{\left[
\partial_\mu\phi \partial^\mu\phi -m^2\phi^2
+l_P^4m^2(\partial_0^2\phi)^2\right]} \ee for which the stress
energy tensor, computed according to (\ref{Tmn}), is%
\be\label{tmnex1}%
T_{\mu\nu}=\partial_\mu\phi\partial_\nu\phi
+l_P^4m^2\delta_\mu^0(\ddot\phi\partial_{0\nu}\phi -\dddot\phi
\partial_\nu\phi)-g_{\mu\nu}{\cal L}. \ee When evaluated for a
plane-wave, and integrated over a sufficiently large box (as done
in Section~\ref{model}), this leads to bulk energy and momentum
\bea {\cal E}&=&T_{00}V={1\over 2} A^2 \omega^2 V(1+2l_P^4m^2\omega^2)\\
{\cal P}_i&=&T_{0i} V={1\over 2} A^2 \omega k_i V(1+2l_P^4m^2\omega^2). \eea %
At once we see that the soccer ball problem has been eliminated.
Consider a case where the quantum particles are definitely
sub-Planckian, with $\omega\ll E_P$ and $m\ll E_P$. Then
regardless of how large $\cal E$ is, we have
\be {\cal E}^2 -{\cal P}^2\approx (Nm)^2 \ee%
with ${\cal E}=N\omega$  and ${\cal P}_i=Nk_i$, and
\be\label{Napprox} N\approx {1\over 2}V A^2\omega .\ee%
This is true even if ${\cal E}\gg E_P$, thus eliminating the
soccer ball problem. Notice that $N$ here does not need to be the
actual number of quanta, a matter to be refined in
Sections~\ref{quantize} and \ref{abnormal}. Interpreting $N$ as
the macroscopic parameter defined as $N={\cal E}/\omega$, the
point is that departures from (\ref{Napprox}) are a function of
$l_P\omega$ and therefore negligible for subPlanckian frequencies.
Regardless of the second quantization details (and what the
actual number of particles is) there is no soccer ball problem.

Bearing laser physics in mind (see Introduction) we consider
another example, where the dispersion relation for a massless
particle is given by: \be\label{disp221} {\omega^2\over
1-(l_P\omega)^2}-{\mathbf k}^2=0. \ee Such a dispersion relation
entails a frequency dependent speed of light (see,
e.g.~\cite{amelvsl,leej}). A varying speed of light has been
considered in a variety of
circumstances~\cite{vslmoff,am,steph,vslrev}, some of which
cosmological in nature. Before applying (\ref{presc}) we rewrite
these relations as \be \omega^2(1+l_P^2\omega^2{\mathbf k}^2)-
{\mathbf k}^2=0 \ee so that we now get field equation \be
(\partial_\mu\partial^\mu -l_P^2 \partial_0^2\partial_i^2)\phi=0.
\ee A Lagrangian giving this equation is \be {\cal L}={1\over
2}{\left[
\partial_\mu\phi \partial^\mu\phi
+l_P^2(\partial_0\partial_i\phi)^2\right]} \ee and so the relevant
components of the stress energy tensor are \be
T_{0\nu}=\dot\phi\partial_\nu\phi
+l_P^2(\partial_{0i}\phi\partial_{i\nu}\phi -\partial_i^2\dot\phi
\partial_\nu\phi)-g_{0\nu}{\cal L}. \ee Evaluated for a
plane-wave, and integrated over a large volume so as to eliminate
the oscillatory terms, we get \bea {\cal E}&=&T_{00}V={1\over 2}
A^2 \omega^2 V{1+(l_P\omega)^2
\over 1-(l_P\omega)^2}\label{E2}\\
{\cal P}_i&=&T_{0i} V={1\over 2} A^2 \omega k_i V{1+(l_P\omega)^2
\over 1-(l_P\omega)^2}\label{P2}
\eea %
and once more we see that there is no soccer ball problem. If
$\omega\ll E_P$, then these expressions reduce to the undeformed
ones regardless of the amplitude, and so anomalous behavior is
triggered by Planckian $\omega$, not by ${\cal E}\sim E_P$.

What happens in these two cases is actually very general. Due to
the linearity of the field equation and the quadratic nature of
the stress energy tensor, ${\cal E}$ and $\cal P$ will always be
quadratic in amplitude, and be deformed by a function of
$l_P\omega$ rather than a function of $l_P {\cal E}$. This
immediately eliminates the soccer ball problem.
This is true for general Lagrangians of the form \be {\cal
L}=-{1\over 2}\phi{\left[\partial_\mu\partial^\mu
\sum c_n (l_P\partial_0)^{2n} +m^2 \right]} \phi,\ee%
of the form
\be {\cal L}={1\over 2}
\partial_\mu\phi \partial^\mu\phi+
\sum c_n {\left((l_P\partial_0)^{n}\partial_i\phi\right)}^2, \ee
or similar generalizations for other Lagrangians considered in the
examples above. It is believed that some realizations associated with
non-commutative geometry~\cite{kappa} can be cast as HOD field theories,
and therefore they will fall under this category.

The fact that these theories naturally bypass the soccer ball
problem doesn't mean that interesting bulk behavior is not present
if the quanta are Planckian, i.e. if $\omega\approx E_p$. Take
expressions (\ref{E2}) and (\ref{P2}) (or Eq.~\ref{number2}). For
quanta with  $\omega\approx E_P$ we see that for the same
amplitude $A$ a wave carries a much larger density of
particles (indeed this is infinite for $\omega = E_P$). The
implication is that the same laser beam intensity can be reached
with a much lower amplitude, or conversely it takes much more
energy to excite a wave with a given amplitude when its color
approaches the Planck frequency.

\section{Second quantization}\label{quantize}
The field $\phi$ in the previous section is a classical field and
its amplitude a $c$-number, not a creation/annilation operator. No
mention of the quantization procedure has been made or is
necessary. All we learn is that a macroscopic object (here a
classical wave) with $\omega\ll E_P$ does not run into Planckian
behavior even if ${\cal E}$ is comparable or larger than $E_P$. By
itself this is a solution of the soccer ball problem; however, if
we want to know in detail how bulk quantities are deformed as a
function of ${\cal E}$ and $N$, and in particular whether the
prescription $l_P\rightarrow l_P/N$ is correct, we need to know
how the number of quanta is defined. Here we present a simple
quantization procedure leading to the prescription in~\cite{leej}.

The main remark is that if we adopt a quantization procedure in
which the wave's energy and momentum satisfies
${\cal E}=N\omega$ and  ${\cal P}_i=N k_i$
(i.e. Eqns. (\ref{nom}) and (\ref{nk}))
then this is equivalent to the solution
$l_P\rightarrow l_P/N$ proposed in~\cite{leej}. Take the first
example in Section~\ref{emwaves}. If (\ref{nom}) and (\ref{nk})
are correct then (\ref{number0}) is deformed as
\be\label{number1} N={1\over 2}V A^2\omega (1+2l_P^4m^2\omega^2) \ee%
and so %
\be {\cal E}^2 -{\cal P}^2=N^2(\omega^2-k^2)=(Nm)^2(1-(l_P \omega)^4) \ee%
which can be rearranged into%
\be {{\cal E}^2 -{\cal P}^2\over 1-{\left(l_P{\cal E}\over
N\right)}^4}=(Nm)^2 .\ee%
By comparing with (\ref{disp11}) we see that the wave, seen as a
collection of particles, satisfies dispersion relations which can
be obtained from the quanta's relations by replacing the mass with
the sum of the quanta's masses and $l_P$ by $l_P/N$, as intuited
in ~\cite{leej}. The same is true of the second example; if Eqns.
(\ref{nom}) and (\ref{nk}) are true then
\be\label{number2} N={1\over 2}V A^2\omega {1+(l_P\omega)^2 \over
1-(l_P\omega)^2} \ee and again we can derive \be {{\cal E}^2 \over
1-{\left(l_P{\cal E}\over N\right)}^2}-{\cal P}^2=0 \ee to be
contrasted with (\ref{disp221}).

This conclusion is quite general and therefore the veracity of the
prescription $l_P\rightarrow l_P/N$ depends on whether we can
implement a second quantization procedure for which Eqns.
(\ref{nom}) and (\ref{nk}) hold true, with $N$ standing for the
actual quantum number operator (which should have an integer
spectrum). For the amplitudes to become proper creation and
annihilation operators they must satisfy%
\be\label{acom}%
[a_{\mathbf k}, a^\dagger_{\mathbf k'}]=\delta_{\mathbf {k k'}}.%
\ee%
Then $N_{\mathbf k}=a^\dagger_{\mathbf k}a_{\mathbf k}$ is indeed
the number operator: it has an integer spectrum and ``counts'' the
number of quanta present in states belonging to a Fock space set
up with $a^\dagger$ as usual. This is merely a definition, its
physical content residing in the expression linking fields and amplitudes%
\be\label{expk}%
\phi=\sum_{\mathbf k}{1\over \sqrt {2 V C(\omega)}}[a_{\mathbf
k}e^{-ik\cdot x} + a^\dagger_{\mathbf
k}e^{ik\cdot x} ]\ee%
or, equivalently, in the ``convenience factor'' $C(\omega)$ in
this expression. In the undeformed theory $C=\omega$, a fact that
follows from canonical quantization, but here we shall leave it as
a free function and then investigate the implications of different
choices.

Specifically we want to know if it is possible for the quantum
Hamiltonian and mometum
to be given by
\bea %
H&=&\sum_{\mathbf k}\omega N_{\mathbf k}\\\label{Hamilt}%
{\hat P}_i&=&\sum_{\mathbf k}k_i N_{\mathbf k},\label{Momen}
\eea%
essentially the required conditions (\ref{nom}) and (\ref{nk}).
Even though there are ambiguities in defining conjugate momenta in
HOD
theories, we can certainly define the quantum Hamiltonian as%
\be%
H=\int d^3x T_{00}%
\ee%
where $T_{00}$ is to be read off from (\ref{tmnex1}) (and
similarly for the momentum using $T_{0i}$). The field $\phi$ in
this expression is to be replaced by its quantum version
(\ref{expk}), under a normal ordering prescription. The result is%
\be%
H=\sum_{\mathbf k}{\omega^2\over C(\omega)}(1+2m^2l_P^4\omega^2)
N_{\mathbf k}%
\ee%
and thus it is possible to realize (\ref{Hamilt}) if we
choose%
\be\label{Com1}%
C(\omega)=\omega(1+2m^2l_P^4\omega^2).%
\ee%
The question is now: to what canonical quantization procedure does
this correspond? One view is that in HOD theories expressions
(\ref{acom}) and (\ref{expk}) are more fundamental than
the canonical quantization postulate%
\be\label{canon}%
[ \phi({\mathbf x}, t),\Pi({\mathbf y},t)]=i\delta({\mathbf
x}-{\mathbf y}),%
\ee%
where $\Pi$ is the momentum conjugate to $\phi$ (not simply
defined for HOD theories). We point out that (\ref{acom}) and
(\ref{expk}) with $C(\omega)$ given by (\ref{Com1}) follow from
canonical quantization (\ref{canon}) for a deformed momentum%
\be \Pi=\dot \phi -2m^2l_P^4\dddot\phi\ee%
Given the ambiguities in defining conjugate momenta for HOD
theories, this proposal is certainly possible, if not unique.
Notice that this calculation is only possible because there is no
upper bound on $|{\mathbf k}|$, so that
we still have the identity%
\be%
\sum_{\mathbf k}e^{i{\mathbf k}\cdot {\mathbf x}}=V\delta({\mathbf
x}).\ee %
In theories where there is a minimal wavelength as well as a
maximum energy, this is no longer true and therefore (\ref{canon})
can never be realized. The view that (\ref{acom}) and (\ref{expk})
are more fundamental to second quantization is probably more
sensible in such cases.

In this example we have worked backwards, but such a construction
is always possible (as long as there is no upper bound on
$|{\mathbf k}|$). For the second example in Section~\ref{emwaves}
the momentum conjugate to $\phi$ should be chosen as%
\be%
\Pi=\partial_0(1-2l_P^2\partial_i^2)\phi\ee%
or even
\be%
\Pi=\partial_0{1+(l_P\partial_0)^2\over 1-(l_P\partial_0)^2}\phi\ee%
to obtain similar results. In either case (\ref{canon}) leads to
(\ref{acom}) and
(\ref{expk}) with%
\be%
C(\omega)=\omega{1-(l_P\omega)^2\over 1+(l_P\omega)^2}\ee%
and it can be checked that (\ref{Hamilt}) is correct, ensuring
(\ref{nom}) and (\ref{nk}). The bulk dispersion relations
therefore satisfy the rule that $l_P$ should be replaced by
$l_P/N$, where the particle number $N$ has now been defined
rigorously.

It is of course possible to follow other second quantization
procedures, and a rather exotic one will be examined in
Section~\ref{abnormal}. As the previous section shows, none of
this affects the fact that HOD theories {\it cannot} have a soccer
ball problem. However, something different but akin to the soccer
ball problem may be found if an exotic second quantization is
chosen, quite different from the one proposed in this section.

\section{Non-linear field theories}\label{NQL}
HOD theories may be criticized on a number of grounds. One may
question the stability of the Cauchy problem or even suspect that
such theories contain ghosts upon suitable field redefinitions.
None of these criticisms has been definitely proved or disproved.
Yet there are ways to introduce deformed dispersion relations into
field theory that don't invoke higher order derivatives. The
conceptual gains are paid for by heavy technical complexity,
because such theories must be necessarily non-linear. It turns out
that these theories are even better protected against the soccer
ball problem than those considered in the previous Sections.

The idea is that $k_\mu$ should roughly correspond to
$\partial_\mu\phi$ when setting up the Lagrangian. For example one
may propose a Lagrangian of the form%
\be%
{\cal L}={1\over 2}{\left[{\partial_\mu\phi\partial^\mu\phi\over
1+(l_P^2\partial_0\phi)^2}-m^2 \phi^2\right]}  \ee%
for theory (\ref{disp1}), or of the form
\be%
{\cal L}={1\over 2}{\left[\partial_\mu\phi\partial^\mu\phi -
m^2\phi^2(1-(l_P^2\partial_0\phi))^4\right]}
\ee%
for theory~(\ref{disp11b}). In general these theories have a
Lagrangian that is not quadratic in the fields. Non-Quadratic
Lagrangians (NQL) introduce non-linearities into the field
equation and therefore one loses the superposition principle.

For definiteness we consider as an example a theory with
Lagrangian
\be%
{\cal L}={1\over 2}{\left[\partial_\mu\phi\partial^\mu\phi -
m^2\phi^2(1-(l_P^2\partial_0\phi))^2\right]}
\ee%
for which the field equation is%
\be\label{KGNQL}%
[\partial_\mu\partial^\mu + m^2]\phi +m^2l_P^4
(\phi\dot\phi^2+\phi^2\ddot\phi)=0. \ee%
Treating the last two terms (let's call them $\delta s$) as a
perturbation, and
trying out a solution of the form%
\be%
\phi=A\cos(\omega t-kx) + \delta\phi%
\ee%
we find that to leading order the extra terms are%
\be%
\delta s=-{1\over 2} m^2l_P^4A^3\omega^2(\cos\Phi + \cos 3\Phi) \ee%
where the phase $\Phi$ is the usual $\omega t - kx$.
From the terms in $\cos\Phi$ we thus find the condition%
\be \label{dispnl}{\omega^2 -{\mathbf k}^2\over 1-{1\over 2}(l_P^2
A\omega)^2}=m^2 \ee that is, we discover that the dispersion
relations are deformed with a controlling parameter of the order
$l_P^2 A\omega$. Thus in these theories the wave's amplitude has a
direct effect on the quanta dispersion relations. This is due to
the non-linearity of the theory.

But there is another effect, due to the $\cos 3\Phi$ term in
$\delta s$. This should be cancelled by a perturbation
$\delta\phi$, also proportional to $\cos 3\Phi$, specifically: %
\be \delta\phi=-{A\over 16}(l_P A\omega)^2\cos((3(\omega t - k
x)). \ee This is an instance of the so-called phenomenon of
ringing: in addition to the fundamental mode, a higher harmonic is
necessarily excited. In this case the higher harmonic has three
times the frequency of the fundamental mode, and its amplitude is
suppressed by a factor of $(l_P A\omega)^2$.

Expression (\ref{dispnl}) is an interesting result. It appears
that in this theory the quanta's dispersion relation knows about
the bulk wave and not just about the frequency of the quanta.
However this is not quite a version of the soccer ball problem,
because the relevant bulk wave parameter is not its energy ${\cal
E}$ but the energy density $T_{00}={\cal E}/V$. Indeed the rough
condition for anomalous behavior is
\be \label{condanol}A^2\omega^2\sim T_{00}\sim {{\cal E}\over V}\sim l_P^{-4}%
\ee i.e. we need the energy {\it density}, rather than the total
energy ${\cal E}$,  to be Planckian. This is also the parameter
controlling deformations of bulk behavior, e.g. ${\cal E}$ and
$\cal P$. So it seems that both quantum and bulk behavior are
controlled by the same parameter, which is in fact a feature of
the bulk wave.

Condition (\ref{condanol}) is even more restrictive than
$\omega\sim E_P$, since we now need both the quanta's energy and
their number density to be Planckian. In general we find that in
NQL theories Planckian effects are more suppressed than in HOD
theories, both in the quantum and bulk dispersion relations.
However we have learnt an interesting lesson, as the next Section
will highlight: {\it it is possible for the quanta's properties to
be controlled by a bulk feature}.

\section{Some orientation}\label{laser}%
In this section we pause to summarize our findings so far. Having
pointed out the experimental potential in the possibility that
Planckian behavior is triggered by ${\cal E}\sim E_P$, we
encountered severe difficulties in realizing this property
in field theory. We investigated DSR
implementations based on HOD Lagrangians:%
\be%
{\cal L}={\cal L}(\phi,\partial_\mu\phi;(l_P\partial_0)^n\phi) \ee
and on NQL%
\be%
{\cal L}={\cal
L}(\phi,\partial_\mu\phi;\phi^2(l_P^2\partial_0\phi)^n).
\ee%
We then distinguished between two levels of phenomenology,
microscopic (relative to the quanta, and to $k_\mu$) and
macroscopic (relative to the bulk wave, and ${\cal E}$ and ${\cal
P}_i$). The first relates to dispersion
relations for the quanta, i.e. the replacement%
\be \omega^2-{\mathbf k}^2=m^2 \rightarrow q(\omega,k,m)=0\ee%
where the deformation function $q$  depends on $l_P$ multiplied by
some energy, not necessarily $\omega$. The latter relates to the
behavior of matter in bulk, investigated here in the extreme case
of a classical wave, made up of coherent quanta. In
general we expect a deformation%
\be {\cal E}^2-{\cal P}_i^2=M^2\rightarrow Q({\cal E},{\cal
P},M)=0\ee%
via a possibly different function $Q$ controlled by $l_P$
multiplied by some other energy, not necessarily $\omega$ or
${\cal E}$.  We saw that in HOD theories%
\bea q&=&q(\omega,k,m;l_P\omega)\nonumber\\
Q&=&Q({\cal E},{\cal P},M;l_P\omega) \eea%
whereas in NQL theories %
\bea q&=&q{\left(\omega,k,m;l_P^4{{\cal E}\over V}\right)}\nonumber\\
Q&=&Q{\left({\cal E},{\cal P},M;l_P^4{{\cal E}\over V}\right)}. \eea%
Thus in HOD theories the Planckian nature of the quanta triggers
both microscopic and macroscopic anomalies, whereas in NQL
theories it is a bulk parameter, the energy density, that controls
anomalous behavior (both microscopic and macroscopic). In either
case the low energy world is well protected from the threat of
Planckian effects.


But in the realm of abstract possibilities two interesting
alternatives might occur. Firstly, anomalous macroscopic behavior
might be triggered by a Planckian bulk energy (${\cal E}\sim
E_P$),  with
\be\label{anom1}%
Q=Q({\cal E},{\cal P},M;l_P{\cal E}). \ee%
This is the soccer-ball problem. NQL theories, while not realizing
this scenario, suggest an even more insidious second possibility.
It could also be that
\be\label{anom2} q=q(\omega,k,m;l_P{\cal E})\ee%
i.e., microscopic QG behavior might be triggered by a Planckian
bulk energy (in the sense ${\cal E}\sim E_P$) where ${\cal E}$ is
the energy of the coherent packet to which the quanta belong. As
pointed out in the introduction, as long as either of these
possibilities happens {\it only} for coherent collections of
particles this is interesting rather than problematic.

The realization of either possibility in classical field theory is
highly unlikely, however. It would require an
integral-differential field
equation, for example:%
\be%
[\partial^\mu\partial_\mu + m^2]\phi -l_Pm^2\phi\int d^3 x
\dot\phi^2=0. \ee%
This equation approximately accepts plane-wave solutions, with dispersion
relations%
\be {\omega^2-{\mathbf k}^2\over 1-l_P{\cal E}}=m^2.%
\ee %
But it would also entail severe non-locality: the value of the
field at one point is directly entangled with the field everywhere
else.

The turn of phrase just used points to an interesting new avenue.
Perhaps the non-locality required to realize (\ref{anom1}) or
(\ref{anom2})  is not present in the classical field theory, but
is instead part of the quantization process, in a suitable
alternative to Section~\ref{quantize}. Quantum entanglement is
non-local, and so is the phase coherence between all the quanta
making up a laser. This may be just the ``non-local'' ingredient
required by  the anomalies (\ref{anom1}) or (\ref{anom2}). This we
investigate in the next section.

\section{Anomalous second quantization}\label{abnormal}
In this Section we search for a quantization procedure that might
lead to (\ref{anom1}) or (\ref{anom2}). Since we seek anomalous
behavior with ${\cal E}\sim E_P$ but $\omega\ll E_P$, we may
assume $\omega\ll E_P$ and ${\cal E}/V\ll E_P^4$. Thus we may as
well investigate anomalous quantization of an undeformed classical
field theory (even if the considerations to follow in effect apply
to HOD and NQL theories, and further anomalies will be present if
$\omega\sim E_P$).

Since the undeformed Klein-Gordon
equation is approximately valid, we have classical plane wave
solutions $\phi=A\cos(\omega t- kx)$ with $k_\mu=\{\omega,k_i\}$
such that, approximately%
\be%
\omega^2-{\mathbf k}^2=m^2\label{om2}.\ee%
The bulk quantities
${\cal E}$ and ${\cal P}_i$ satisfy:%
\bea%
{\cal E}&=&{V\over 2} A^2\omega^2\\
{\cal P}_i&=&{V\over 2} A^2\omega k_i\\
{\cal E}^2-{\cal P}^2&=&(Nm)^2=M^2 \eea%
with $N$ {\it formally}  defined as before. All of this is purely
classical.

In setting up second quantization in Section~\ref{quantize} we
adopted a deformation of the usual factor $C(\omega)=\omega$
(c.f. Eqn.~(\ref{expk})), but
the deformation was taken to be a function of the
frequency. For all we know it could be a function of $l_P{\cal E}$
where ${\cal E}$ is the energy of the coherent packet under
examination. Then,
\be\label{expk1}%
\phi=\sum_{\mathbf k}\sqrt{\hbar(l_P{\cal E)}\over 2 V
\omega}[a_{\mathbf k}e^{-ik\cdot x} + a^\dagger_{\mathbf
k}e^{ik\cdot x} ]\ee%
with
\be%
[a_{\mathbf k}, a^\dagger_{\mathbf k'}]=\delta_{\mathbf {k k'}}.%
\ee%
Once more $N_{\mathbf k}=a^\dagger_{\mathbf
k}a_{\mathbf k}$ properly counts the number of quanta present in
the states of the standard Fock space, set up with $a^\dagger$
as usual.

This construction is equivalent to an ${\cal
E}$-dependent Planck constant (explaining the notation used). It
follows from canonical quantization with
\be\label{canon1}%
[ \phi({\mathbf x}, t),\dot\phi({\mathbf y},t)]=i\hbar(l_P{\cal
E})\delta({\mathbf
x}-{\mathbf y}).%
\ee%
Mimicking the calculation performed in Section~\ref{quantize}
we find that
\bea\label{Hamilt1} %
H&=&\sum_{\mathbf k}\hbar(l_P{\cal E})\omega N_{\mathbf k}\\
{\cal P}_i&=&\sum_{\mathbf k}\hbar(l_P{\cal E})k_i N_{\mathbf k}. %
\eea%
Thus additivity is preserved in the form:
\bea {\cal E}&=&NE\\
{\cal P}_i&=& Np_i \eea%
but now we find that the energy-momentum $p_\mu=(E,-{\mathbf p})$ of
the quanta is non-trivially related to their frequency and wavenumber:
\be\label{defnew}%
p_\mu=\hbar(l_P{\cal E}) k_\mu.%
\ee%
We have so far ignored the distinction between the quanta's energy
$E$ and their frequency $\omega$. We are now zooming in on this
distinction, and implementing the anomalous element there.

But the dispersion relations for $k_\mu$ are undeformed (or rather,
they're deformed according to $l_P\omega$ and we are taking the
limit $l_P\omega\ll 1$). Thus the proposed deformation of the
relation between $p_\mu$ and $k_\mu$ implies that the quanta's
mass-shell condition must be deformed as a function of $l_P{\cal
E}$. For example if%
\be%
\hbar={1\over 1-l_P{\cal E}} \ee%
so that
\bea %
E&=&{\omega\over 1-l_P{\cal E}}\\
p_i&=&{k_i\over 1-l_P{\cal E}} \eea%
then (\ref{om2}) leads us to the quantum mass-shell conditions %
\be%
{E^2-p^2\over (1-l_P{\cal E})^2}=m^2. \ee%
Therefore if we consider a high intensity laser, and perform
experiments sensitive to the energy, rather than the frequency of
its quanta (for example, photo-electric type experiments), we
should become sensitive to QG. The most dramatic implication
follows from the proved additivity (\ref{Hamilt1}), which implies
\be%
N={1\over 2}A^2\omega V (1-l_P{\cal E}). \ee%
This creates a version of the soccer ball problem: we cannot build
a coherent wave packet with an energy larger than $E_P$.  As
${\cal E}$ approaches $E_P$ the packet behaves more and more like
a single quantum, and the energy of a single quantum (for which
now $E\approx {\cal E}$) cannot exceed $E_P$. In this limit we
have blurred the distinction between bulk and quantum, thus
creating a version of the soccer ball problem for coherent matter.

This construction has parallels with~\cite{sab}, but also striking
differences. In that work it was proposed that%
\be\label{defsab}%
p_i=k_i \hbar (l_P k).
\ee%
leading to a deformed Heisenberg uncertainty principle and a
frequency dependent Planck's constant. As in~\cite{sab} we
postulate that the relation between $p_\mu$ and $k_\mu$ is
deformed, but as a function of $l_P{\cal E}$ rather than $l_P k$.
Furthermore in~\cite{sab} the quantum mass-shell relations are
assumed to remain undeformed, implying deformed dispersion
relations for $k_\mu$. We have exactly reversed this assumption.
This is because HOD and NQL theories teach us that deformations of
(\ref{om2}) can never be a function of $l_P\cal E$ (the purpose of
this paper).

To restate the case in a more physical language, in~\cite{sab} it
takes more and more energy to produce a quantum as its wavelength
approaches $l_P$. Here we suggested that it may take more and more
energy to make a quantum (of whatever wavelength, not necessarily
small), as the energy of the coherent packet to which it belongs
approaches ${\cal E}_P$. Then $N\approx 1$ and the distinction
between classical and quantum disappears. The packet behaves like
a single quantum particle for which $E_P$ cannot be exceeded.

\section{Conclusions}
In DSR Planckian behavior is triggered whenever the energy of
elementary particles approaches $E_P$. That Planckian behavior may
also set in when  the total energy in a collection is close to
$E_P$ constitutes the ``soccer ball problem''. In this paper we
pointed out that there is no obvious contradiction with experiment
if only {\it coherent} superpositions of particles with collective
energy close to $E_P$ display Planckian behavior. The proverbial
``soccer balls'' aren't coherent and the strongest lasers have
bulk energies well below $E_P$.

Unfortunately, upon closer inspection, we found that in classical
field theory realizations of DSR it is very difficult, if not
impossible, {\it not} to solve the soccer ball problem. Our
analysis was not exhaustive -- for example we left out
non-commutative field theory -- but the results obtained are more
than suggestive. We conclude that the soccer ball concern is a red
herring, a mental block peculiar to simplistic kinematic models.
With the community having spent so long worrying about the
``problem'', it is ironic that now when a ``good side'' is found,
difficulties are encountered in realizing it at all.

But we found a better experimental window of opportunity. Even
though classical waves with sub-Planckian frequencies can never
display anomalous bulk properties, it is possible for their quanta
to behave unusually. We proposed that in sets of coherent
particles the relation between the quanta's energy and frequency,
and between their momentum and wavenumber, is deformed whenever
the energy of the coherent packet nears $E_P$. In other words
Planck's constant is deformed as a function of $l_P{\cal E}$ and,
like most deformations considered in DSR, diverges for ${\cal
E}\rightarrow E_P$. Then, in the same way that in DSR there cannot
be elementary particles with energy bigger than $E_P$, it becomes
impossible to construct a coherent packet with energy larger than
$E_P$. As the packet energy gets larger each quantum absorbs more
and more energy (for fixed frequency), until for ${\cal
E}\rightarrow E_P$ the whole packet effectively becomes a single
quantum, with sub-Planckian frequency, but Planckian energy. The
distinction between bulk and particle then disappears, and it is
impossible to push the energy of particle or packet beyond $E_P$.
It is therefore impossible to have a Laser deliver more than about
a TeraJoule of coherent light. The implied limitations to military
systems, such as those inspired by SDI, are so tragic they
shouldn't even be contemplated.

Once regarded as an embarrassment the soccer ball ``problem''
never exists in its classical field theory realization. But
Pandora's box is opened with the realization that an important
test, rather than a paradox, follows for a version of the
phenomenon that only affects coherent collections of particles. We
found that it is possible (but not generic) for the quantum
mass-shell condition to be deformed as a function of $l_P{\cal
E}$, where ${\cal E}$ is the total energy of a coherent wave
packet. The consequent implications of high intensity Laser
projects for QG phenomenology are highlighted for the first time
in this paper.

{\bf Acknowledgements} I'd like to thank Giovanni Amelino-Camelia,
Kim Baskerville, Sabine Hossenfelder, Shahn Majid and Lee Smolin
for very helpful comments on various versions of this manuscript.

\label{lastpage}

\end{document}